\documentclass[prl,twocolumn,showpacs]{revtex4}
\usepackage{amsfonts,amsmath,mathrsfs,epsfig,amsbsy,bm,verbatim}

\begin{document}

\title{Majorana fermions in carbon nanotubes}
\author{Jay D. Sau$^1$}

\author{Sumanta Tewari$^{2}$}
\affiliation{$^1$Department of Physics, Harvard University, Cambridge, MA 02138 \\
$^2$Department of Physics and Astronomy, Clemson University, Clemson, SC
29634}

\begin{abstract}
We show that carbon nanotubes (CNT) are good candidates for
realizing one-dimensional topological superconductivity with Majorana fermions localized near the end points. The physics behind topological superconductivity in CNT
is novel and is mediated by a recently reported curvature-induced spin-orbit coupling which itself has a topological origin.
In addition to the spin-orbit coupling, an important new requirement for a robust topological state is broken chirality symmetry about the nanotube axis.
 We use topological arguments
  to show that, for recently realized strengths of spin-orbit coupling and
 broken chirality symmetry, a robust topological 
 gap $\sim 500$ mK is achievable in carbon nanotubes.

\end{abstract}

\maketitle

\paragraph{Introduction:} Topological superconductors (TS) in dimensions $D=2,1$ with broken time-reversal (TR) symmetry have
recently attracted a lot of attention \cite{Read-Green,sau-et-al,long-PRB,roman,oreg,Tewari-NJP,zhang-tewari,Choy,Martin,Flensberg}. These systems support Majorana fermion excitations at order parameter defects such as vortices and sample
edges. Majorana fermion
excitations, with second quantized operators $\gamma$ satisfying the self-hermitian condition $\gamma^\dagger = \gamma$, can be construed as quantum particles which
are their own antiparticles \cite{Wilczek-3}. The self-hermiticity of Majorana fermions (MF) leads to a $2D$ quasiparticle exchange statistics which is non-Abelian \cite{Read-Green,Nayak-Wilczek}.
The non-Abelian statistics of MF's can be used as a robust quantum mechanical resource to implement fault-tolerant topological quantum computation (TQC) \cite{Kitaev,nayak_RevModPhys'08}.

In 1D TS with broken TR symmetry, Majorana fermion modes are supposed to be trapped at the two ends of a quantum wire \cite{long-PRB,roman,oreg}, which,
 in a $2D$ quantum wire network \cite{Alicea}, can potentially lead to successful demonstration of non-Abelian statistics as well as TQC
 \cite{Alicea,Sau-Universal,Beenakker}. Recently, a number of systems with effective Rashba-type spin-orbit coupling and broken TR-invariance have been proposed to be in an
  appropriate TS state with localized MF modes \cite{sau-et-al,long-PRB,roman,oreg,Tewari-NJP,zhang-tewari,Choy,Martin,Flensberg}. In this paper we add to this list an attractive new candidate - carbon nanotubes (CNT) - which have
  already become important experimental systems due to their remarkable electronic and transport properties \cite{Roche-RMP,Mceuen,Flensberg2}. Note that the superconducting proximity effect from an $s$-wave superconductor on carbon nanotubes has also been realized \cite{CNTproximity}.
 By employing robust
   topological arguments 
  we show that CNTs in proximity to ordinary superconductors such as Al, Nb present a
  good platform for realizing Majorana fermions at the ends of the nanotube provided the chirality symmetry about the tube axis can be
   broken. Using parameters from
   recent experiments \cite{Mceuen,Flensberg2}  we show that, under appropriate external conditions, it is possible to realize a
  TS state gap protecting the Majorana fermion excitations as high as $\sim 500$ mK.

  The physics behind the topological state in CNT is novel and is mediated by a spin-orbit coupling (SOC) which itself has a
  topological origin. SOC in carbon-based systems (such as Graphene) is generally negligible due to a small ($\sim 8$ meV)
  atomic SOC of carbon atoms which gets further suppressed by the band-structure. Nonetheless, CNT, which is cylindrically rolled-up Graphene, can have a reasonably strong SOC due
  to curvature-induced topological effects \cite{Mceuen,Flensberg2,Ando,Brataas}. The band structure of CNT inherits the two momentum space Dirac cones of Graphene, but the momenta along
the circumferential direction are discrete and quantized due to finite size effects. Electronic states with quantized circumferential momenta arising from different
 Dirac cones have opposite chirality
(indexed by chirality index $\sigma_z = 1, -1$ in Eq.~(\ref{eq:Hamiltonian1})), which
indicates the clockwise or anticlockwise circulation about the nanotube axis. In one circulation around
the circumference the electronic orbitals undergo a complete rotation about spin, which results in an effective
topological spin-orbit coupling with strength inversely proportional to the tube diameter $d$ \cite{Mceuen,Flensberg2,Ando,Brataas}. It has been shown recently that for tubes with diameter $d = 5$ nm, the spin-orbit
energy splitting $\Delta_{SO}$ can be as high as $\sim 0.4$ meV. Note that $\Delta_{SO}$ arises from a topological Berry phase accumulated by
orbital rotation around the circumference and has no explicit momentum dependence as is customary in Rashba or Dresselhaus couplings in ordinary
semiconductors. We show below that this type of spin-orbit coupling, even in the absence of an explicit momentum dependence, can produce a one-dimensional TS state provided time-reversal and chiral symmetries are also broken in CNT.

 The topological spin-orbit coupling and broken TR invariance (due to an external transverse magnetic field) are not sufficient by themselves
to realize a finite gap in the TS state in the nanotube. One more crucial ingredient for a non-zero TS state gap is broken chiral symmetry about
the tube axis. In the presence of chiral symmetry (i.e., rotational symmetry about the tube axis) all
states in the nanotube can be classified as circulating clockwise or anticlockwise about the tube axis (i.e., $\sigma_z$ in Eq.~(\ref{eq:Hamiltonian1}) is a good quantum number). In one of the main results of this paper we show on quite general grounds that in this case the gap in the topological state, even if the topological invariant is non-trivial, identically vanishes. Therefore, for a finite TS state gap with
robust Majorana fermions,
the nanotube Hamiltonian must mix and split the chiral eigenstates. Such couplings between the clockwise and the anticlockwise states can be induced 
by breaking the rotational symmetry about the nanotube axis using the superconducting proximity effect itself or it can arise from disorder (which can also break this symmetry).
 In a recent paper
Jespersen et al. has shown \cite{Brataas} the existence of such a term $\Delta_{K,K'}$ in a multi-electron CNT with a magnitude of $\Delta_{K,K'}$ of the same order as $\Delta_{SO}$. By calculating the appropriate topological invariant (the Pfaffian invariant \cite{Kitaev-1D, Ghosh}) and the bulk quasiparticle gap we show that, in such a nanotube, a finite TS state gap $\sim 500$ mK is achievable with a modest value $\sim 1$ T of an external magnetic field. Our calculations and results establish the single walled carbon nanotube as an attractive new candidate for topological physics and Majorana fermions potentially leading to TQC.

\paragraph{Hamiltonian for carbon nanotube and topological spin-orbit coupling:}
The states in a carbon nanotube are characterized by a chirality index $\sigma_z=\pm 1$, a pseudo-spin $\rho_z$
and a parallel spin component $s_z$. The Hamiltonian of a CNT in the presence of a transverse magnetic field $B_{\perp}$,
but no chirality breaking term, is given by,
\begin{equation}
H_{CNT}=v_F\sigma_z\{(k_{||}+K_{||}\sigma_z)\rho_x+(k_0+\alpha \sigma_z s_z)\rho_y\} +B_{\perp}s_x
\label{eq:Hamiltonian1}
\end{equation}
where the Hamiltonian, for $B_{\perp}=0$, commutes with the anti-unitary time-reversal operator $T=i s_y \sigma_x K$ which satisfies $T^2=-1$. Here $\rho_{x,y}$ are Pauli matrices in the sub-lattice degree of freedom, while $\sigma_z$ is the Pauli
matrix whose eigenstates arise from different valleys $K,K'$ of the underlying Graphene sheet.  The wave-vector $k_{||}$ is the wave-vector
of the electron along the axis of the carbon nanotube. Since the Dirac cones $K,K'$ of the underlying Graphene sheet
occur at wave-vectors with opposite signs, they have opposite projection $K_{||}\sigma_z=\pm K_{||}$ along the
 axis of the nanotube. Confinement along the circumference of the nanotube leads to a minimum circumferential
momentum of magnitude $k_0$ and a corresponding minimum gap $v_F k_0$ in the Dirac cone where $v_F$ is the fermi velocity of the underlying Graphene sheet. Electrons near the bottom
 of each band near $K$ or $K'$ have circumferential group velocities in the clockwise or in the anti-clockwise
direction (corresponding to $\sigma_z=\pm 1$) respectively. The atomic spin-orbit coupling of carbon leads to electrons
with different spin orientations to experience different Aharonov-Bohm fluxes, shifting $k_0\rightarrow k_0+\alpha s_z\sigma_z$,
where $\alpha$ is the topological spin-orbit coupling strength \cite{Mceuen}. Finally
 $B_{\perp}$ is the Zeeman splitting of the electrons from an applied magnetic field in the transverse ($x$-) direction. For the moment
 we are ignoring any chirality breaking term in the Hamiltonian (so $\sigma_z$ is a good quantum number). This is because
  we want to prove below one of the central results of this paper which states that, in the presence of chirality symmetry,  there cannot be a non-zero TS state gap even when the Pfaffian topological invariant given by Eq.~(\ref{eq:q}) is non-trivial. It follows that, for a robust TS state in CNT, mixing of the clockwise and the anti-clockwise states in the nanotube is a crucial requirement as is a non-zero spin-orbit coupling $\alpha$.

\paragraph{BdG Hamiltonian and the Pfaffian invariant:} In the presence of an $s$-wave superconductor proximity coupled to the nanotube, a superconducting
pair potential $\Delta$ will be induced on the CNT.
In this case, the singlet BdG Hamiltonian for the CNT that anti-commutes with the particle-hole operator $\Lambda=-i \tau_y T$
is given by,
\begin{eqnarray}
H_{BdG}&=&\Big[v_F(k_{||}\sigma_z\rho_x+K_{||}\rho_x+k_0 \sigma_z\rho_y+\alpha s_z \rho_y)-\mu\Big]\tau_z \nonumber\\&+&B_{\perp}s_x+\Delta \tau_x.
\label{eq:BdG1}
\end{eqnarray}
The BdG Hamiltonian for the CNT is topologically non-trivial and supports Majorana fermions at the ends provided
the Pfaffian invariant  written as,
\begin{equation}
Q=\textrm{sgn}\Big(Pf(\Lambda H_{BdG}(k_{||}=0))\Big)\times\Big(Pf(\Lambda H_{BdG}(k_{||}=\pi)\Big)\label{eq:q}
\end{equation}
is $-1$ \cite{Ghosh}. However, in addition to the non-trivial value of $Q$, for a useful topological state with localized Majorana fermions protected from thermal decoherence, the CNT must have a non-zero (proximity-induced) topological superconducting gap. Below we show that, for $Q=-1$, a non-zero TS state gap is not possible as long as the chirality (given by the $\sigma_z$ in Eq.~(\ref{eq:Hamiltonian1})) is a good quantum number.

\paragraph{Absence of topological gap in the presence of chiral symmetry:}

The BdG Hamiltonian in Eq.~(\ref{eq:BdG1}) commutes with $\sigma_z$ so that different chirality sectors $\sigma_z=\pm 1$ are
decoupled. Here we show that such Hamiltonians are necessarily gapless whenever it has a non-trivial Pfaffian
invariant. To do this, we first assume that $Q=-1$ so that the Pfaffians of $H_{BdG}$ at the particle-hole symmetric points $k_{||}=0,\pi$ have
opposite signs. Next construct another BdG Hamiltonian,
\begin{eqnarray}
\tilde{H}_{BdG}&=&\Big[v_F(k_{||}\rho_x+K_{||}\rho_x+k_0\sigma_z\rho_y+\alpha s_z\rho_y)-\mu\Big]\tau_z\nonumber\\ &+&B_{\perp}s_x+\Delta \tau_x,
\label{eq:BdG2}
\end{eqnarray}
such that $\tilde{H}_{BdG}(k_{||})$ is particle-hole symmetric at \emph{each value} $k_{||}$ (i.e., not just at $k_{||}=0,\pm \pi$ as in Eq.~(\ref{eq:BdG1})). This allows us to define a Pfaffian for $\tilde{H}_{BdG}$ at each value of $k_{||}$ and not just at $k_{||}=0,\pm \pi$.  Note also that $\tilde{H}_{BdG}(k_{||})$ has the same spectrum as
$H_{BdG}(k_{||})$ and, in addition, $\tilde{H}_{BdG(k_{||}=0)}=H_{BdG(k_{||}=0)}$ and the equality also holds for $k_{||}=\pm \pi$.
Therefore, if $Q=-1$ for $H_{BdG}$, so is $Q$ for $\tilde{H}_{BdG}$.
Since we have assumed that $H_{BdG}$ does indeed have $Q=-1$, it follows that the Pfaffians of $\tilde{H}_{BdG}$ at $k_{||}=0,\pi$ also
have opposite signs.
Therefore the Pfaffian of $\tilde{H}_{BdG}$ must vanish at some $k_{||}$ between $k_{||}=0$ and $k_{||}=\pi$. Note that since $\tilde{H}_{BdG}$
 is particle-hole symmetric at \emph{all} points $k_{||}$ between $0$ and $\pi$ (but $H_{BdG}$ is not) we can define a
  Pfaffian for $\tilde{H}_{BdG}$ at arbitrary $k_{||}$ and it must be zero somewhere on the $k_{||}$ line for $Q=-1$. The Pfaffian of a BdG
  Hamiltonian can vanish at a $k_{||}$ point only when one of the eigenvalues vanish at that point. Further, since $\sigma_z$ commutes with both $H_{BdG}$ and
$\tilde{H}_{BdG}$, all eigenstates have definite values of $\sigma_z$. Therefore, if $\tilde{H}_{BdG}$ is gapless at
some value of $k_{||}$ and $\sigma_z$, so is $H_{BdG}$. This shows that carbon nanotubes are gapless at some value of $k_{||}$ whenever the Pfaffian invariant is non-trivial, $Q=-1$.

\paragraph{Chiral symmetry breaking and nanotube band-structure:} To obtain a non-zero topological gap when the Pfaffian invariant
$Q=-1$ one must add a chirality breaking term
to the nanotube Hamiltonian which does not commute with $\sigma_z$.
Such a term can arise from
disorder effects and/or broken rotational symmetry about the nanotube axis due to the substrate. Note that the superconducting proximity effect
 itself can be used to break the rotational symmetry about the CNT axis. With broken rotational symmetry about the
 tube axis the clockwise and the anticlockwise states are no longer eigenstates of the one-electron Hamiltonian and their splitting can be
   represented by the term $\Delta_{K,K'}$. Such a term, with a magnitude comparable to the energy splitting due to the topological spin-orbit coupling, has recently been experimentally observed \cite{Flensberg2}.  Adding this term to Eq.~(\ref{eq:Hamiltonian1}) the nanotube Hamiltonian changes to,
\begin{eqnarray}
H_{CNT}&=&v_F\sigma_z\{(k_{||}+K_{||}\sigma_z)\rho_x+(k_0+\alpha \sigma_z s_z)\rho_y\} +B_{\perp}s_x\nonumber\\ &+&\Delta_{K,K'}\sigma_x.
\label{eq:Hamiltonian2}
\end{eqnarray}

\begin{figure}
\centering
\includegraphics[scale=0.3,angle=0]{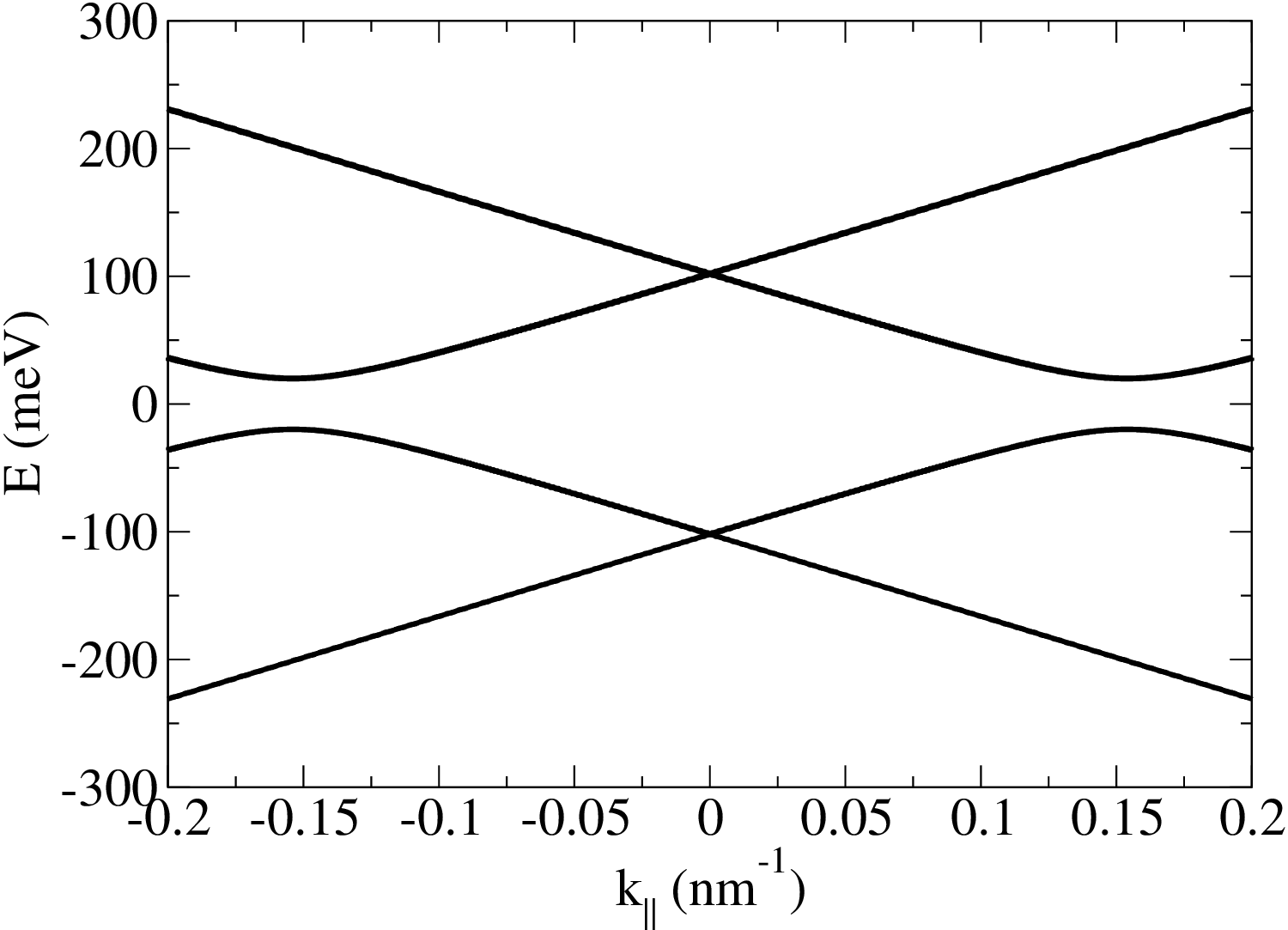}
\caption{Bandstructure of a (9,3) chiral CNT. Such CNTs have a band minimum away from $k_{||}=0$ and therefore the
bands have a linear intersection near $k_{||}=0$.}
\label{fig:band1}
\end{figure}

\begin{figure}
\centering
\includegraphics[scale=0.3,angle=0]{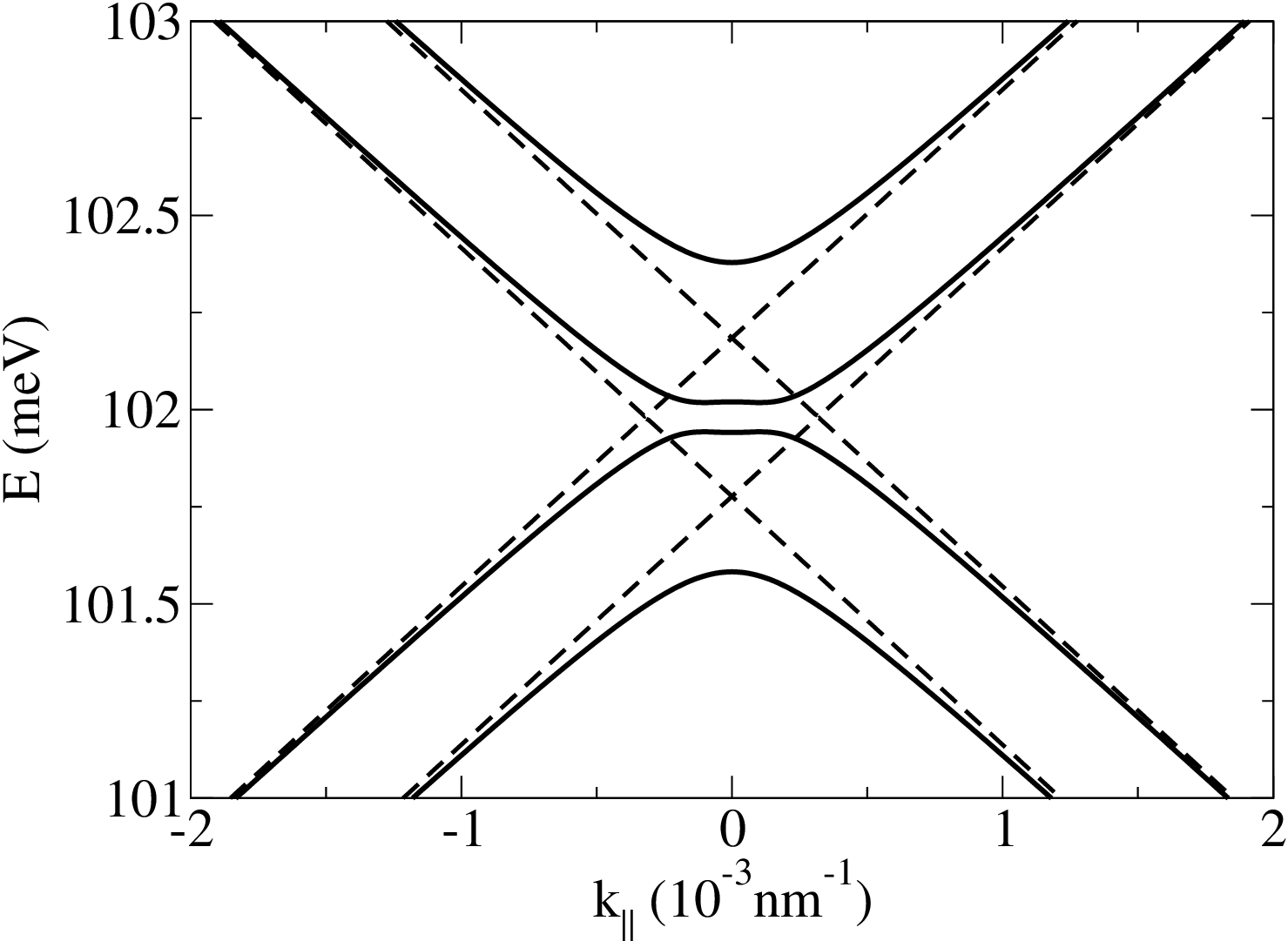}
\caption{The combination of $\Delta_{KK'}$ opens up a gap in the dashed band-structure
with an odd number of fermi surfaces. We have chosen $\alpha=400 \mu$eV,$B_{\perp}=400  \mu$eV (corresponding to $0.5$ T) and $\Delta_{KK'}=200 \mu$eV.}
\label{fig:band2}
\end{figure}

To understand the effects of $\Delta_{K,K'}$ let us first ignore its contribution to the band structure. The dispersion of the conducting states in the nanotube are then given by
\begin{equation}
\epsilon(k,\sigma,s)=v_F\sqrt{(k_{||}-K_{||}\sigma_z)^2+(k_0+\alpha s\sigma_z)^2}-\mu,
\end{equation}
where we have used the fact that the Hamiltonian commutes with $\sigma_z$.
For simplicity, we consider a nanotube in the semiconducting regime so that we can focus only on the states in the conduction band. The corresponding
band-structure near the relevant intersection point at $k=0$ is plotted in Fig.~\ref{fig:band1}. Each curve in Fig.~\ref{fig:band1} actually represents two bands (because of near spin-degeneracy) split by much smaller energy scales $B_{\perp}$ and $\alpha$,
 and these are shown as dashed lines in Fig.~\ref{fig:band2}. 

From the dashed lines in Fig.~\ref{fig:band2} it is clear that even though the conduction bands ($E>0$) are split by the topological
spin-orbit coupling $\alpha$ and the transverse field $B_{\perp}$, all values of the chemical potential give rise to an even
number of Fermi surfaces. We know, on the other hand, that for a topological superconductor with an odd number of Majorana fermions at each end
of the nanotube it is necessary to obtain an odd number of bands at $k_{||}=0$ \cite{Kitaev-1D}. The solid curves in Fig.~\ref{fig:band2} depict
the band-structure near $k_{||}=0$ with a non-zero chirality breaking term $\Delta_{K,K'}$ in the nanotube Hamiltonian. The term $\Delta_{K,K'}$
opens up a gap in the one-electron spectrum, but, in contrast to the Rashba-coupled systems, chemical potential in the gap at $k_{||}=0$ does not produce an odd number of Fermi surfaces. There are in fact two ranges in $\mu$ (for the parameters in Fig.~\ref{fig:band2} on both sides of $E \sim 102$ meV),
for which the nanotube has an odd number of Fermi surfaces and, consequently, is in a gapped topological state in the presence of a proximity-induced pair potential $\Delta$.

\paragraph{Topological state and Majorana fermions with chiral symmetry breaking:}  
The BdG Hamiltonian of the nanotube including the topological spin-orbit coupling $\alpha$, transverse field $B_{\perp}$, and the chirality breaking term $\Delta_{KK'}$ can be written as,

\begin{eqnarray}
H_{BdG}&=&\Big[v_F(k_{||}\sigma_z\rho_x+K_{||}\rho_x+k_0\sigma_z\rho_y+\alpha s_z \rho_y)\nonumber\\&+&\Delta_{K,K'}\sigma_x-\mu\Big]\tau_z +B_{\perp}s_x+\Delta \tau_x.
\label{eq:BdGchiral}
\end{eqnarray}

\begin{figure}
\centering
\includegraphics[scale=0.3,angle=0]{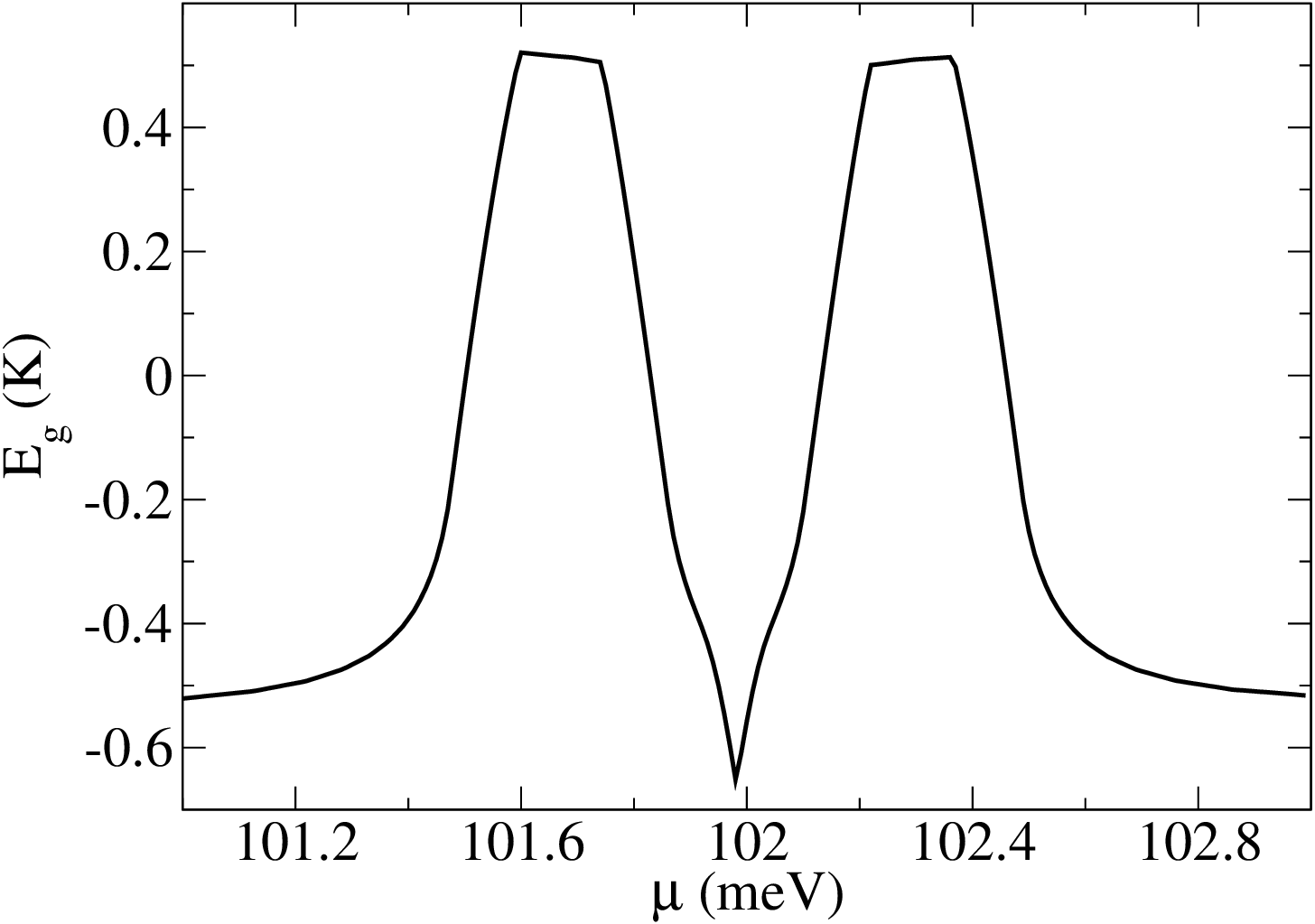}
\caption{Topological superconducting gap in carbon nanotube as a function of chemical potential $\mu$. The gap is multiplied by the negative of the topological invariant $Q$ (Eq.~\ref{eq:q}),
so it is positive in the nanotube TS state with Majorana fermions ($Q=-1$) and negative in the topologically trivial ($Q=1$) superconducting state. We have chosen a proximity induced $s$-wave pair potential
$\Delta=300 \mu$eV.}
\label{fig:gap}
\end{figure}

For the Hamiltonian in Eq.~\ref{eq:BdGchiral} we first calculate the $\mathbb{Z}_2$ invariant $Q$ given in Eq.~\ref{eq:q}. In Fig.~\ref{fig:gap}, we plot
the superconducting gap (calculated by diagonalizing the BdG Hamiltonian in Eq.~\ref{eq:BdGchiral}) multiplied by the negative of the $\mathbb{Z}_2$ invariant $Q$ as
a function of the chemical potential $\mu$. Since $Q=-1 (1)$ signals the topologically non-trivial (trivial) superconducting state, positive values
of this product indicate the TS state in Fig.~\ref{fig:gap}. It is clear that there are well defined ranges of the chemical potential in which the
nanotube supports Majorana fermions with a non-zero TS state gap $\sim 500$ mK.  

\paragraph{Conclusion:} In this paper we show that carbon nanotubes are an attractive candidate for realizing 
one-dimensional time-reversal breaking topological superconducting state with zero-energy Majorana fermions localized near the ends.
The physics of the TS state in CNTs is mediated by a novel, curvature-induced, spin-orbit coupling which is topological in itself.
We show that, despite the presence of this coupling (reported in recent experiments \cite{Mceuen,Flensberg2} to be substantial, $\Delta_{SO}\sim 0.4$ meV), there
cannot be a non-zero TS state gap in the nanotube in the presence of chirality symmetry \emph{even in a parameter regime where the relevant
topological invariant is non-trivial}. Thus, to obtain a robust topological state in CNTs with end-state Majorana fermions, the chiral symmetry about the tube axis must be broken. This can be achieved by the superconducting proximity effect itself or by disorder which may also break
the rotational symmetry about the nanotube axis. Using recently reported values of $\Delta_{SO}$ and the chiral
symmetry breaking term $\Delta_{K,K'}$ we show that in carbon nanotubes a robust topological gap $\sim 500$ mK is achievable.

J.S. thanks the Harvard Quantum Optics Center for support. S.T. thanks DARPA and NSF for support.


\end{document}